\documentclass[12pt]{article}

\def\##1{\underline #1}
\def\=#1{\underline{\underline #1}}
\def\4#1{\underline{\underline{\underline{\underline #1}}}}

\def\.{\mbox{ \tiny{$^\bullet$} }}

\def\les{\left[}
\def\ris{\right]}

\def\ric{\right\}}

\def\c#1{\cite{#1}}

\def\curl{\nabla\times}
\def\Re{{\rm Re}}

\def\eps{\epsilon}
\def\epso{\eps_0}
\def\muo{\mu_0}
\def\ko{k_0}
\def\etao{\eta_0}
\def\epsr{\eps_r}
\def\mur{\mu_r}

\def\tde{\tan\delta_\eps}
\def\tdm{\tan\delta_\mu}
\def\nr{n'}
\def\ni{n''}

\date{}

\begin{document}

\baselineskip .582cm   

\begin{center}
\vskip 0.2cm
{\large\bf AN ELECTROMAGNETIC TRINITY FROM \\
``NEGATIVE PERMITTIVITY" AND \\
``NEGATIVE PERMEABILITY"}\\[20pt]
{\bf Akhlesh Lakhtakia}
\\[20pt]
{\em CATMAS -- Computational \& Theoretical Materials Sciences Group\\
Department of Engineering Science and Mechanics\\
Pennsylvania State University, University Park, PA 16802--6812, USA}\\[25pt]
\end{center}

\begin{abstract}
An electromagnetic trinity comprising vacuum, anti--vacuum, and nihility is 
postulated~---~after making use of materials with ``negative permittivity" and ``negative
permeability"~---~to illuminate the structure of electromagnetic theory, at least
insofar as the relationship of phase velocity with Poynting vector is concerned.
\\

\end{abstract}

\section{Introduction}
Illuminating normally one of the two relevant faces of a wedge by a
10~GHz beam, Shelby {\em et al.\/} \c{SSS} observed
that the angle of emergence of the beam from the other face changed
sign when the teflon wedge was replaced by one made of the
so--called left--handed material (LHM) \c{SPVNS, SK}.
Assuming that their {\em LHM\/} is homogeneous and isotropic, and
virtually disregarding dissipation, they concluded that it possesses a negative
index of refraction.

Never mind that the so--called {\em LHM\/} does not 
actually possess handedness \c{Bel}, and therefore is mis--named.
Never mind that this material is actually a particulate composite \c{SSS}
containing  particles that are not electrically small~---~especially
when the fiber--glass content of the LHM is factored in~---~and therefore
is not homogenizable into a medium with local response properties \c{MLW}.
Never mind also that the {\em LHM\/} is actually anisotropic and must
have some scattering losses. 

A negative index of refraction
is not needed {\em per se\/}, because all that it connotes is that
the phase velocity is directed opposite to the velocity of energy
transport \c{Chen}~---~as shown in the
Appendix. But the phase velocity is an inessential
concept \c{BH}, whether in vacuum or in materials. 
What really matter are the solutions of Maxwell
equations that propagate energy on a desired trajectory,
which is chosen epistemically.

Anyhow, virtually all publications on {\em LHMs\/} 
that I have come across
\c{SSS}--\c{SK},  \c{P}--\c{LK} deal substantively only with non--dissipative
materials, with just one exception \c{LK}. Perhaps, in time to come,
samples of these materials with proven negligible dissipation at some
frequency will emerge. In the meanwhile, they permit me to postulate an
electromagnetic trinity comprising (i) vacuum, (ii) anti--vacuum, and (iii) nihility.
This trinity illuminates the structure of electromagnetic theory, at least
insofar as the relationship of phase velocity with Poynting vector is concerned.

\section{Vacuum}
Vacuum, of course, is well--known. Also called {\em free space\/},
it is a matter--free medium~---~the substrate on which microscopic
electromagnetics is formulated and later homogenized for macroscopic
research \c{Jac,Rob}. The frequency--domain constitutive relations of vacuum
are specified as
\begin{equation}
\left. \begin{array}{l}
{\bf D}(x,y,z,\omega) = \epso\, {\bf E}\,(x,y,z,\omega)\\[10pt]
{\bf B}(x,y,z,\omega) = \muo \, {\bf H}(x,y,z,\omega)
\end{array}\ric\,,
\end{equation}
where $x$, $y$ and $z$ are the cartesian components
of the position vector; $\omega$ is the angular frequency; $\epso=8.854\times 10^{-
12}$~F~m$^{-1}$ and
$\muo=4\pi\times 10^{-7}$~H~m$^{-1}$. The phase velocity and the
wavenumber
of a plane wave in vacuum are co--parallel.

\section{Anti--vacuum}
The constitutive relations of anti--vacuum are postulated as
\begin{equation}
\left. \begin{array}{l}
{\bf D}(x,y,z,\omega) = -\epso \,{\bf E}(x,y,z,\omega)\\[10pt]
{\bf B}(x,y,z,\omega) = -\muo \,{\bf H}(x,y,z,\omega)
\end{array}\ric\,,
\end{equation}
Among the more significant properties of anti--vacuum is that the
phase velocity of a plane wave therein is opposite in direction to the wavevector.
Hence, anti--vacuum is complementary to vacuum.

Clearly, anti--vacuum does not exist, but it could be simulated
at a particular spatial frequency as follows:
Suppose one could fabricate a {\em LHM\/} whose relative permittivity scalar
$\epsr(\omega)$ is purely real--valued at a certain angular frequency $\tilde\omega$,
and whose relative permittivity scalar $\mur(\omega)$ is such that
$\mur(\tilde\omega) = \epsr(\tilde\omega)$.
Let electrically small spheres of this {\em LHM\/} be homogeneously
but randomly dispersed in vacuum. As both mediums would be impedance--matched,
the effective constitutive parameters of 
the particulate composite thus formed would have to satisfy the relation
$\mur^{eff}(\tilde\omega) =\epsr^{eff}(\tilde\omega)$
\c{L96}.
Then, $\epsr^{eff}(\tilde\omega) = -1$ according to the Bruggeman
formalism \c{L96},  provided the
volume fraction  of the {\em LHM\/} is
\begin{equation}
f_{LHM} = \frac{2}{3}\, \frac{\epsr(\tilde\omega) -2}{\epsr(\tilde\omega)-1}\,, \quad 0 \leq f_{LHM}\leq 1\,.
\end{equation}
As an example, $f_{LHM}= 0.8$ if $\epsr(\tilde\omega) = -4$. Hence,
this composite would mimic anti--vacuum at $\omega=\tilde\omega$.

\section{Nihility}
Nihility is the electromagnetic nilpotent, with the following
constitutive relations
\begin{equation}
\left. \begin{array}{l}
{\bf D}(x,y,z,\omega) = {\bf 0} \\[10pt]
{\bf B}(x,y,z,\omega) = {\bf 0}
\end{array}\ric\,.
\end{equation}
Wave propagation cannot occur in nihility, because $\curl {\bf E}(x,y,z,\omega) ={\bf 0}$
and $\curl {\bf H}(x,y,z,\omega)  = {\bf 0}$ in the absence of sources therein \c{K71, L92}. The directionality of the phase velocity relative to the
wavevector in nihility is thus a non--issue.

The Bruggeman formalism is unable to predict the realization of nihility
as a particulate composite either
by mixing anti--vacuum and vacuum, or by mixing
an  {\em LHM\/} with an isoimpedant commonplace material (such as teflon). This
is because the polarizability and the magnetizability
(per unit volume) of an isotropic dielectric--magnetic sphere  embedded
in nihility do not depend on the constitutive parameters of the medium
that the sphere is made of. 

The Maxwell Garnett formalism \c{L96}, however, predicts
that a homogeneous dispersion of electrically small, anti--vacuum spheres in vacuum
is effectively equivalent to nihility, provided the volume fraction of
anti--vacuum is $0.25$. 
Conversely, if matter were scooped out of anti--vacuum in the form
of electrically small spheres, the resulting Swiss--cheese composite would
also mimic nihility, if the volume fraction of anti--vacuum were $0.75$.
Thus, nihility too could be mimicked by  particulate
composites at $\omega=\tilde\omega$.

\section{Concluding Remarks}
Whereas vacuum and anti--vacuum are mutually complementary in
regard to the directionality of the phase velocity with respect to
the Poynting vector, that issue does not arise for nihility because
waves cannot propagate inside it.
Of the three, vacuum alone is matter--free. Both
anti--vacuum and nihility would have to be simulated~---~at a specific angular
frequency~---~as particulate composites which would ultimately require the
fabrication of virtually non--dissipative {\em LHMs\/} that are impedance--matched
to vacuum. When would this become possible at all, let alone
routinely, is anybody's guess.

\small{

}

\section*{Appendix}
Let us consider an isotropic, dielectric--magnetic, homogeneous
medium with relative permittivity $\epsr = -a(1-i\tde)$ and 
relative permittivity $\mur=-b(1-i\tdm)$ at a certain angular frequency
$\omega$.  Let $a $,
$b $, $m = +\sqrt{ab}$, $\tde $ and $\tdm $ all be positive real, so that
$\epsr$ and $\mur$ are in accord with the Lorentz model
\c{SSS, OS}, with $\omega$ significantly larger than the resonant angular frequencies
of the medium. 
The $\exp(-i\omega t)$ time--dependence is implicit.
To ensure that dissipation is moderate, let $\tde < 0.1$ and $\tdm < 0.1$.

The electromagnetic phasors of a planewave traveling along
the $+z$ axis in the chosen medium are given by
\begin{equation}
\left.
\begin{array}{ll}{\bf E}(z,\omega) = A\,{\bf u}_x \exp (i\ko n z)\\[10pt]
{\bf H}(z,\omega) = \frac{n}{\mur} \,\frac{A}{\etao}\, {\bf u}_y \exp (i\ko n z)
\end{array}\ric\,,
\end{equation}
where $A$ is a complex--valued amplitude,
$\ko$ is the  vacuum wavenumber, $\etao$ is the intrinsic
impedance of vacuum, and $n^2 = \epsr\mur$.
Writing $n=\nr+i\ni$, we see that  the time--averaged 
Poynting
vector is given as
\begin{equation}
\label{Py}
{\bf P}(z,\omega) =
{\bf u}_z\,\Re\{\frac{n}{\mur}\} \frac{\vert A\vert^2}{2\etao}\,\exp(-2\ko\ni z)
= {\bf u}_z \,P_z(z,\omega)\,,
\end{equation}
the velocity of energy transport is co--parallel with ${\bf P}(z,\omega)$,
and
the phase velocity $v_{ph} = \omega/\ko\nr$.
Correct to
first order in both $\tde$ and $\tdm$, we get $n=\pm m \les 1-i(\tde+\tdm)/2\ris$
and $\Re\{\frac{n}{\mur}\} = \mp m/b$.

For energy to flow along the $+z$ axis, we must have $P_z>0$.
Furthermore, the directions of energy attenuation and energy flow must coincide
(as befits a passive medium), so that $\ni>0$. Both of these
criterions lead to $\nr<0$,
which means that the phase velocity is pointed exactly opposite to the
common direction
of energy attenuation and flow.

\end{document}